# Grassmann representation in qubit informatics and superlogic

Valery V. Smirnov [*]

*Faculty of Physics, Saint Petersburg State University, Russia*

The Grassmann representation for the system of qubits, is considered. The treatment is based on natural description of the qubits system as fermions and uses coherent states of fermions. The quantum logic gates are represented in two forms – by symbols of operations and by partial differential operators on the symbols of the states. The considered representation of quantum logic is called as superlogic. The examples are given for classical logic operations of negation, conjunction, disjunction and for reversible three-bit Toffoli gate and its quantum generalization – three-qbit Deutsch gate. The representation for composite gates is also considered. Path integral represenatation in Grassmann algebra for quantum automaton with qubits memory is described. In particular for the autonomous automaton corresponding to the special case of general dynamic system this description differ from path integral representation in commutative memory phase space by the specific nonliner term in superaction.

## 1. Introduction

Natural description of quantum systems with a set of binary (two levels) cells (qubits) is based on Fermi operators. Conventional treatment of fermions systems is based on Grassmann representation. One of the most straightforward approach to this topic is based on the coherent states formalism. Coherent states are known to be a valuable tool in quantum systems description. Fermi operators are Heisenberg-Weyl supergroup generators and defines corresponding set of generalized coherent states in terms of [1]. Coherent states of fermions were introduced in the works of Martin [2] and further studied in [3–5]. It should be noted that coherent states of fermions doesn't exist in original quantum states space and need its extension by means of Grassmann algebra. Nevertheless these states can be used in calculations and gives correct results in compact form. They are parametrized by points in classical phase space for fermions which are generators of Grassmann algebra with involution.

Coherent states provide symbols of operators and of vectors in quantum states space [1, 6]. These symbols and operations with them give another representation for corresponding quantum system. In particular we have another representation for qubits and operations with them. Corresponding representation of quantum logic (Q-logic) may be called as superlogic (S-logic). Superlogic is adequate mathematical formalism for quantum logic gates [7–11]. Grassmann representation may be also used for path integral description of quantum automaton behaviour.

Let us make some notes on used terminology.

The term superlogic but in somewhat different meaning is used in work [12]. The authors introduces Boolean supermanifold using conventional constructions in a supermathematics [13]. The reduced manifold of Boolean supermanifold has the structure of Boolean manifold [14] with topology of a Boole algebra of propositions defined by considering the implication as an order relation. The superlogic in terms of [12] gives extension of classical Boolean logic (B-logic) but its relation to Q-logic needs additional consideration. Our treatment of superlogic in contrast is based on explicit and exact representation of Q-logic and gives corresponding extension of B-logic. Note that the supermanifold in our approach is complex and has trivial body. However in more general quantum systems with Bose and Fermi degrees of freedom the supermainfold should have dimensions of both parities.

---
[*] E-mail: valery_smirnov@mail.ru

Note also that the term Q-logic in some works [15] has the meaning which differ from used in this paper.

## 2. Grassmann representation

Some technical details of Grassmann representation, used notations and terminology conventions are given for convenience in Appendix section.

Suppose we have a system of $N$ qbit cells. Binary basis of it states will be denoted as $B_N$ and corresponding quantum qbit states space as $F_N$. Any operator on $F_N$ can be represented as polynomial over creation $a_k^+$ and annihilation $a_k^-$ ($k = 1,\ldots,N$) operations which satisfy canonic anticommutation relations /A.1/ [6]. So qbit cells system can be treated as fermions system.

Rather popular is the representation of fermions system in Grassmann algebra [6]. Convenient technical tool here are fermions coherent states. Coherent states can be used for definition of symbols of the states and of the operators. Symbols are the elements of Grassmann algebra $\Lambda_{2N}$ with generators /A.2, A.3/ and gives corresponding representation of the states and operators on $\Lambda_{2N}$. Generators /A.2, A.3/ are often treated as variables of classical phase space of fermions system. From this point of view symbols can be treated as functions on phase space and realize quantum to classical correspondence principle. Note that symbols of the states $F_N$ are functions on only half of generators /A.3/ (in classical terms they are treated as variables of classical configuration space) and consequently are the elements of subalgebra $\Lambda_N^*$ in Grassmann algebra $\Lambda_{2N}$. This leads to another realization of operators on $F_N$ by means of partial differential operators on $\Lambda_N^*$ – so called Fock-Bargmann representation.

Any B-logic function (gate) $f: B_N \to B_M$ has linear continuation within Q-logic $f: F_N \to F_M$. Important examples are unary operation of negation $not: B_1 \to B_1$ and binary operations $and, or, xor, nand: B_2 \to B_1$.

As an illustration let us consider some of commonly used B-logic operators and corresponding S-logic operators in both representations – by symbols on $\Lambda_{2N}$ and by partial differential operators on $\Lambda_N^*$.

Negation operation has the form
$$not = a^+ + a^-.$$
Its covariant symbol is
$$not(\alpha^*, \beta) = \alpha^* + \beta$$
and Fock-Bargmann representation is
$$not\left(\alpha^*, \frac{\partial}{\partial \alpha^*}\right) = \alpha^* + \frac{\partial}{\partial \alpha^*}.$$

As an example of binary operations let us consider conjunction and disjunction operations. They can be represented by means of basis projectors as
$$and = |0\rangle(\langle 0,0| + \langle 1,0| + \langle 0,1|) + |1\rangle\langle 1,1|,$$
$$or = |0\rangle\langle 0,0| + |1\rangle(\langle 1,0| + \langle 0,1| + \langle 1,1|).$$
Covariant symbols of these operations are undefined because input and output have different dimensions (see Appendix). The matrix symbols are
$$\langle \alpha^*|and|\beta_1, \beta_2\rangle = \langle \alpha^*|0\rangle\langle 0,0|\beta_1, \beta_2\rangle + \langle \alpha^*|0\rangle\langle 1,0|\beta_1, \beta_2\rangle + \langle \alpha^*|0\rangle\langle 0,1|\beta_1, \beta_2\rangle + \langle \alpha^*|1\rangle\langle 1,1|\beta_1, \beta_2\rangle$$
or

$$\langle \alpha^* | and | \beta_1, \beta_2 \rangle = 1 + \beta_1 + \beta_2 + \alpha^* \beta_1 \beta_2$$

and

$$\langle \alpha^* | or | \beta_1, \beta_2 \rangle = \langle \alpha^* | 0 \rangle \langle 0, 0 | \beta_1, \beta_2 \rangle + \langle \alpha^* | 1 \rangle \langle 1, 0 | \beta_1, \beta_2 \rangle + \langle \alpha^* | 1 \rangle \langle 0, 1 | \beta_1, \beta_2 \rangle + \langle \alpha^* | 1 \rangle \langle 1, 1 | \beta_1, \beta_2 \rangle$$

or

$$\langle \alpha^* | and | \beta_1, \beta_2 \rangle = 1 + \alpha^* (\beta_1 + \beta_2 + \beta_1 \beta_2).$$

Fock-Bargmann representations are

$$and\left(\alpha^*, \beta^*, \frac{\partial}{\partial \beta^*}\right) = \frac{\partial}{\partial \beta_1^*} \frac{\partial}{\partial \beta_2^*} \left(\alpha^* + \beta_2^* \beta_1^*\right) + \frac{\partial}{\partial \beta_1^*} + \frac{\partial}{\partial \beta_2^*},$$

$$or\left(\alpha^*, \beta^*, \frac{\partial}{\partial \beta^*}\right) = \frac{\partial}{\partial \beta_1^*} \frac{\partial}{\partial \beta_2^*} \left(\alpha^* + \beta_2^* \beta_1^*\right) + \alpha^* \left(\frac{\partial}{\partial \beta_1^*} + \frac{\partial}{\partial \beta_2^*}\right).$$

For some reasons reversible logical gates can be of interest [9]. In this case input and output variables have equal dimensions $N = M$. Let us consider, for example, so called controlled-controlled-not (cc_not, or Toffoli) three-bit gate $cc\_not: B_3 \to B_3$. It can be represented by the identity operator 1 and projectors on basis as

$$cc\_not = 1 + (|1, 1, 1\rangle - |1, 1, 0\rangle)(\langle 0, 1, 1| - \langle 1, 1, 1|).$$

Its symbol is

$$cc\_not(\alpha^*, \beta) = 1 + \alpha_3^* \alpha_2^* (\alpha_1^* - 1)(1 - \beta_1) \beta_2 \beta_3$$

and Fock-Bargmann representation is

$$cc\_not\left(\alpha^*, \frac{\partial}{\partial \alpha^*}\right) = 1 + \alpha_3^* \alpha_2^* (\alpha_1^* - 1)\left(1 - \frac{\partial}{\partial \alpha_1^*}\right) \frac{\partial}{\partial \alpha_2^*} \frac{\partial}{\partial \alpha_3^*}.$$

Quantum generalization of Toffoli three-bit gate is the set of three-qbit Deutsch gates [9] $D(\varphi): F_3 \to F_3$. It can be represented by projectors on basis as

$$D(\varphi) = 1 + |1, 1, 0\rangle ((i \cos(\varphi) - 1)\langle 0, 1, 1| + \sin(\varphi)\langle 1, 1, 1|) + |1, 1, 1\rangle (\sin(\varphi)\langle 0, 1, 1| + (i \cos(\varphi) - 1)\langle 1, 1, 1|),$$

where $\varphi$ is real parameter. Its symbol is

$$D(\varphi)(\alpha^*, \beta) = 1 + \alpha_3^* \alpha_2^* ((i \cos(\varphi) - 1)(1 + \alpha_1^* \beta_1) + \sin(\varphi)(\alpha_1^* + \beta_1)) \beta_2 \beta_3.$$

It can be also represented in exponential form which will be used later in Subsection 3.2

/1/ $\quad D(\varphi)(\alpha^*, \beta) = \exp(H(\varphi)(\alpha^*, \beta)),$

/2/ $\quad H(\varphi)(\alpha^*, \beta) = \alpha_3^* \alpha_2^* ((i \cos(\varphi) - 1)(1 + \alpha_1^* \beta_1) + \sin(\varphi)(\alpha_1^* + \beta_1)) \beta_2 \beta_3.$

The Fock-Bargmann representation is

$$D(\varphi)\left(\alpha^*, \frac{\partial}{\partial \alpha^*}\right) = 1 + \alpha_3^* \alpha_2^* \left((i \cos(\varphi) - 1)\left(1 + \alpha_1^* \frac{\partial}{\partial \alpha_1^*}\right) + \sin(\varphi)\left(\alpha_1^* + \frac{\partial}{\partial \alpha_1^*}\right)\right) \frac{\partial}{\partial \alpha_2^*} \frac{\partial}{\partial \alpha_3^*}.$$

Consider the representation of composite logical gates. In particular the circuits with elementary gates with two inputs and one output have significant practical interest. Suppose $A$ and $B$ are such elementary gates and let us compose two composite gates with tree inputs and one output

$$C1|n_1, n_2, n_3\rangle = A(|n_1\rangle \otimes B|n_2, n_3\rangle),$$

$$C2|n_1, n_2, n_3\rangle = A(B|n_1, n_2\rangle \otimes |n_3\rangle).$$

Using resolution of the identity operator we can find the values of composite operators on the coherent states

$$C1|\beta_1,\beta_2,\beta_3\rangle = A(|\beta_1\rangle \otimes B|\beta_2,\beta_3\rangle),$$
$$C2|\beta_1,\beta_2,\beta_3\rangle = A(B|\beta_1,\beta_2\rangle \otimes |\beta_3\rangle),$$

and inserting resolution of the identity operator /A.6/ before operator $B$ we have following formulas for the matrix symbols

$$\langle\alpha^*|C1|\beta_1,\beta_2,\beta_3\rangle = \int \langle\alpha^*|A|\beta_1,\gamma\rangle\langle\gamma^*|B|\beta_2,\beta_3\rangle \exp(-\gamma^*\gamma)d\gamma^*d\gamma,$$
$$\langle\alpha^*|C2|\beta_1,\beta_2,\beta_3\rangle = \int \langle\alpha^*|A(\gamma)\rangle\langle\gamma^*|B|\beta_1,\beta_2\rangle\otimes|\beta_3\rangle)\exp(-\gamma^*\gamma)d\gamma^*d\gamma.$$

With commutation rule /A.4/ the matrix symbols of $C2$ can be rewritten in form

$$\langle\alpha^*|C2|\beta_1,\beta_2,\beta_3\rangle = \int \left(\langle\alpha^*|A|\gamma,\beta_3\rangle\langle\gamma^*|B|\beta_1,\beta_2\rangle^e + \langle\alpha^*|A|\gamma,-\beta_3\rangle\langle\gamma^*|B|\beta_1,\beta_2\rangle^o\right)\exp(-\gamma^*\gamma)d\gamma^*d\gamma,$$

where for arbitrary Grassmann algebra elements $x = x^e + x^o$ is the expansion on even and odd parts.

Generalizing let us consider the gate $A$ with $N$ inputs and one output and suppose that $k$-th input is replaced with elementary gate $B$ with two inputs and one output. This results in composite gate $C$ with $N+1$ inputs and one output. Let us denote the coherent state of $N+1$ cells as

$$|\beta\rangle = |\beta_1,\ldots,\beta_{N+1}\rangle = |\beta_1,\ldots,\beta_{k-1}\rangle|\beta_k,\beta_{k+1}\rangle|\beta_{k+2},\ldots,\beta_{N+1}\rangle,$$

where brackets repetition means tensor product. For shorter notation we write it as

$$|\beta\rangle = |\beta',\beta'',\beta'''\rangle = |\beta'\rangle|\beta''\rangle|\beta'''\rangle,$$

where $\beta''$ is in $k$-th position (for $k = N$ the last variables $\beta'''$ are omitted). The matrix symbol of composite gate can be written in following form

$$\langle\alpha^*|C|\beta\rangle = \int \left(\langle\alpha^*|A|\beta',\gamma,\beta'''\rangle\langle\gamma^*|B|\beta''\rangle^e + \langle\alpha^*|A|\beta',\gamma,-\beta'''\rangle\langle\gamma^*|B|\beta''\rangle^o\right)\exp(-\gamma^*\gamma)d\gamma^*d\gamma.$$

The matrix symbol of any composite gate can be obtained by repeating of this formula.

Fock-Bargmann representation can be more suitable for representation of composite gates. The composite gate operator is a composition of partial differential operators in Grassmann algebra, for example

$$C1\left(\alpha^*,\beta^*,\frac{\partial}{\partial\beta^*}\right) = A\left(\alpha^*,\beta_1^*,\frac{\partial}{\partial\beta_1^*},\gamma^*,\frac{\partial}{\partial\gamma^*}\right)B\left(\gamma^*,\beta_2^*,\frac{\partial}{\partial\beta_2^*},\beta_3^*,\frac{\partial}{\partial\beta_3^*}\right),$$

$$C2\left(\alpha^*,\beta^*,\frac{\partial}{\partial\beta^*}\right) = A\left(\alpha^*,\gamma^*,\frac{\partial}{\partial\gamma^*},\beta_3^*,\frac{\partial}{\partial\beta_3^*}\right)B\left(\gamma^*,\beta_1^*,\frac{\partial}{\partial\beta_1^*},\beta_2^*,\frac{\partial}{\partial\beta_2^*}\right).$$

For the composition of two logical gates with equal number of inputs and outputs the composite gate symbol is the convolution /A.9/ of the original gates symbols. In Fock-Bargmann representation we have the composition of corresponding partial differential operators.

## 3. Path integral representation of automaton behavior

Algebraic theory describes automaton as representation of an input semigroup in a transformations semigroup of automaton set of states [16]. In case of words input semigroup this corresponds to Moore automaton (without output), in case of time input semigroup – to autonomous automaton. This approach is valid in discrete and in continual cases. In this paper we consider quantum automaton. Formaly quantum automaton corresponds to unitary representation.

It is well known that semigroup representation or for autonomous automaton in other terms evolution operator may be represented in path integral form. In case of probabilty autonomous automaton path integral is of Wiener type and in quantum case it is of Feynman type. Such representations for evolution operator is widely used for example in physics [6].

Quantum automaton with qubits memory is of especial interest. Natural path integral describtion of such systems leads to path integral in Grassmann algebra. General form of path integral representation in Grassmann algebra was found in [17, 18]. In this paper it is applied to quantum automaton with qubits memory. We give explicit path integral representation for autonomous and Moore automaton.

### *3.1 Autonomous automaton*

Autonomous automaton is special case of general dynamic system. Covariant symbol of evolution operator between time moments $t', t''$ has continual path integral representation

/3/ $$U(\alpha^*, \beta) = \int \exp(iS) d\gamma^* d\gamma.$$

Eq. /3/ assumes integration over all virtual trajectories $(\gamma^*, \gamma)$ in phase space $\Lambda_{2N}$ with boundary conditions $\gamma^*(t'') = \alpha^*$, $\gamma(t') = \beta$. Superaction is

/4/ $$S = \int \left( d\gamma^*(t) \cdot \gamma(t) - iH(\gamma^*(t), \gamma(t)) dt - O dt \right),$$

where the term $O$ according to [16] is

/5/ $$O = H^o(\gamma^*(t), \gamma(t)) \int_{t'}^{t} H^o(\gamma^*(s), \gamma(s)) ds$$

and $H^o$ is the odd part in Grassmann algebra of covariant symbol (Hamilton function) $H$ of evolution operator generator.

Without the term Eq. /5/ the expression Eq. /4/ has usual form of classical Hamilton-Kahler action [6]. It is valid for even Hamilton function symbol. The term Eq. /5/ nonlinear for the odd part of Hamiltonian is the specific feature of path integral representation in Grassmann algebra in general case of Hamiltonian symbol with arbitrary Grassmann parity [17]. This follows from the relation

$$\exp(x) \cdot \exp(y) = \exp(x + y + x^o y^o)$$

for arbitrary Grassmann algebra elements.

### *3.2 Moore automaton*

Let $W$ be a concatenation semigroup of input words for Moore automaton, $End\, Q$ – a transformations semigroup of automaton set of states $Q$ and

$$U : W \to End\, Q$$

semigroup representation describing automaton.

For every input word partition

$$w = \sum_{k=0}^{n-1} \Delta w_k, \qquad w, \Delta w_k \in W$$

according to the semigroup property we can write

/6/ $$U(w) = \prod_{k=0}^{n-1} U(\Delta w_k).$$

Suppose that the set of states $Q$ is the quantum system and has the structure of fermions system. In particular it can be the set of binary cells considered above. Then the Grassamann representation can be used for the automaton description.

According to correspondence principal /A.9/ the symbol of the operator /6/ can be written as the product of convolutions

/7/ $$U(w)(\alpha_n^*, \alpha_0) = \mathop{\circ}_{k=0}^{n-1} U(\Delta w_k)(\alpha_{k+1}^*, \alpha_k).$$

Using integral form of the convolution /A.7/ we have another representation in the form of path integral

$$U(w)(\alpha^*, \beta) = \int \prod_{k=0}^{n-1} U(\Delta w_k)(\gamma_{k+1}^*, \gamma_k) \exp(iS') d\gamma^* d\gamma,$$

where restricted action is defined as

$$S' = \sum_{k=0}^{n-1} \Delta \gamma_k^* \gamma_k, \quad \Delta \gamma_k^* = (\gamma_{k+1} - \gamma_k)^*$$

and integration is over all discrete virtual trajectories $(\gamma^*, \gamma)$ in phase space $\Lambda_{2N}$ with boundary conditions $\gamma_n^* = \alpha^*$, $\gamma_0 = \beta$.

The representation /7/ can be useful when there exist an alphabet $A$ of the semigroup of input words $W$ such that for each character $\Delta w \in A$ the symbol $U(\Delta w)(\alpha^*, \beta)$ is known. In particular, when the symbol of transformation operator for the character can be represented in exponential form $U(\Delta w)(\alpha^*, \beta) = \exp(H(\Delta w)(\alpha^*, \beta))$ the symbol for the whole word can be represented in following path integral form

$$U(w)(\alpha^*, \beta) = \int \exp(S) d\gamma^* d\gamma$$

where

$$S = \sum_{k=0}^{n-1} \left( \Delta \gamma_k^* \cdot \gamma_k - H(\Delta w_k)(\gamma_{k+1}^*, \gamma_k) - O_k \right),$$

$$O_k = H^o(\Delta w_k)(\gamma_{k+1}^*, \gamma_k) \sum_{j=0}^{k-1} H^o(\Delta w_j)(\gamma_{j+1}^*, \gamma_j)$$

with boundary conditions $\gamma_n^* = \alpha^*$, $\gamma_0 = \beta$.

A simple example gives Moore automaton based on three-qbit Deutsch gates /1/, /2/ $U(\Delta w) = D(\varphi)$ with concatenation semigroup of input words freely generated by the parameters $\Delta w = \varphi$ treated as an alphabet. Note that this semigroup is not equal to the additive semigroup of real numbers. However for closely connected three-qbit gates $D'(\varphi) = J \cdot D(\varphi)$, where $J = 1 \oplus 1 \oplus -i1$ and 1 is the identity in two dimensions we have

$$D'(\varphi' + \varphi) = D'(\varphi') \cdot D'(\varphi)$$

and it can be defined Moore automaton for the additive semigroup of real numbers $W = R$. But the defined Q-logic gates are B-logic gates only in the trivial case for the parameters $\varphi$ satisfying $D'(\varphi) = 1$.

## 4. Conclusion

Grassmann representation for qubits systems gives their natural description. Corresponding superlogic extending classical and quantum logic can be introduced. Logical gates can be realized in two forms either by means of symbols of operators which are the elements of Grassmann algebra or by means of partial differential operators on the symbols of the states which are the elements of Grassmann algebra with half number of generators – so called Fock-Bargmann representation.

Respectively the composite gates can be represented either by means of integral operations on the symbols or by means of a composition of partial differential operators in Fock-Bargmann representation.

The behaviour of quantum automaton with qubits memory can be described by path integral in Grassmann algebra. In case of autonomous automaton this description differ from path integral representation in commutative memory phase space by the specific nonliner term in superaction. Grassmann representation leads to simple rules of symbolic calculations for qubits systems.

## 5. Appendix

In classical logic (Boolean or shortly B-logic) a bit is a binary cell with two states $B = \{|0\rangle, |1\rangle\}$. In quantum logic (shortly Q-logic) qbit states are the elements of complex linear algebra $F(B)$ with basis $B$. The states of the system of $N$ bits and qbits are correspondingly $B_N = B^N$ and $F_N = F(B)^{N\otimes}$ (power of tensor product $\otimes$). The basis of $N$ qbits can be written as $B_N = B^{N\otimes}$, so any basis element has the form

$$|n\rangle = |n_1\rangle \otimes \ldots \otimes |n_N\rangle \quad \text{(or shortly } |n\rangle = |n_1, \ldots, n_N\rangle\text{),}$$

where $n$ is the function of the cell number with two values $n: \{1, \ldots, N\} \to \{0, 1\}$.

Within usual convention we also consider dual spaces $F_N^+$ with corresponding duality relations. In particular the elements of the dual basis $B_N^+$ are $\langle n| = \langle n_N| \otimes \ldots \otimes \langle n_1|$ (or shortly $\langle n| = \langle n_1, \ldots, n_N|$) and $\langle n|m\rangle = \delta_{n,m}$. The bijection $F_N \xleftrightarrow{+} F_N^+$ induced by relations $\langle n|^+ = |n\rangle$, $|n\rangle^+ = \langle n|$ and $i^+ = i^*$ ($*$ – is complex conjugate) sets nondegenerate Hermitian scalar product on both spaces. This leads to definition of Hermitian adjoint linear operators on both spaces.

The system of qbits allows the description in the form of a fermions system.

First consider one qbit cell.

Let us introduce creation $a^+$ and annihilation $a^-$ operations as

$$a^+|0\rangle = |1\rangle, \quad a^+|1\rangle = 0,$$
$$a^-|0\rangle = 0, \quad a^-|1\rangle = |0\rangle.$$

These operations are Hermitian conjugate pair and satisfy canonic anticommutation relations

$$\{a^+, a^-\} = 1, \quad \{a^+, a^+\} = 0, \quad \{a^-, a^-\} = 0.$$

This allows to treat a qbit cell as a fermions system.

Corresponding operations on dual space are

$$\langle 0|a^- = \langle 1|, \quad \langle 1|a^- = 0,$$
$$\langle 0|a^+ = 0, \quad \langle 1|a^+ = \langle 0|.$$

These operations also satisfy canonic anticommutation relations.

Further we also need the operator corresponding to the third Pauli matrix

$$I = \begin{pmatrix} 1 & 0 \\ 0 & -1 \end{pmatrix},$$

which satisfies the relations

$$I^2 = 1, \quad \{I, a^\pm\} = 0.$$

Now consider the system of qbit cells. Let us introduce creation $a_k^+$ and annihilation $a_k^-$ operations as

$$a_k^\pm = 1 \otimes \ldots \otimes 1 \otimes a^\pm \otimes I \ldots \otimes I,$$

where operators $a^\pm$ are in *k*-th position. These operators satisfy canonic anticommutation relations

/A.1/ $\quad \{a_i^+, a_j^-\} = 1\delta_{i,j}, \quad \{a_i^+, a_j^+\} = 0, \quad \{a_i^-, a_j^-\} = 0.$

Corresponding operators on dual space are

$$\tilde{a}_k^\pm = I \otimes \ldots \otimes I \otimes a^\pm \otimes 1 \ldots \otimes 1,$$

where $a^\pm$ are in $N-k$-th position.

This leads to the possibility to treat a system of qbit cells and a dual system as a fermions systems. Rather convenient is conventional representation of fermions system in Grassmann algebra [6].

Suppose $\Lambda_{2N}$ is Grassmann algebra with $2N$ generators ($2^{2N}$ – is its dimension) and with involution *. Any $N$ generators

/A.2/ $\quad \alpha = (\alpha_1, \ldots, \alpha_N)$

generates subalgebra $\Lambda_N$ in Grassmann algebra $\Lambda_{2N}$. Also suppose that /A.2/ and the set

/A.3/ $\quad \alpha^* = (\alpha_1^*, \ldots, \alpha_N^*)$

generates the whole $\Lambda_{2N}$.

Fermions coherent states are parametrized by $\Lambda_{2N}$ generators and can be formally defined as for bosons system

$$|\alpha\rangle = \exp(\alpha \cdot a^+)|0\rangle,$$

Hereinafter the following notation is used $\quad \alpha \cdot a^+ = \sum_k \alpha_k \cdot a_k^+.$

Fermions coherent states are the element of the extension of $F_N$ with $\Lambda_{2N}$

$$\Phi_N = \Lambda_{2N} \otimes F_N.$$

and are the common eigenstates of annihilation operators with Grassmann eigenvalues

$$a_k^-|\alpha\rangle = |\alpha\rangle\alpha_k.$$

Note that the order of the multipliers is important due to noncommutativity of $\Lambda_{2N}$ and for any $g \in \Lambda_{2N}$

/A.4/ $\quad g\left|(-1)^{p(g)}\alpha\right\rangle = |\alpha\rangle g,$

where $p(g)$ is the parity of $g$ (0 and 1 for even and odd element respectively). In particular for any generator we have $\gamma|-\alpha\rangle = |\alpha\rangle\gamma$.

In the dual space we have adjoint relations

$$\langle\alpha^*| = \langle 0|\exp(a^- \cdot \alpha^*),$$

$$\langle\alpha^*|a_k^+ = \alpha_k^*\langle\alpha^*|,$$

$$|\alpha\rangle^+ = \langle\alpha^*|.$$

The important relations are normalization

$$\langle\alpha^*|\beta\rangle = \exp(\alpha^* \cdot \beta),$$

/A.5/ $\quad \langle\alpha^*|0\rangle = \langle 0|\beta\rangle = 1,$

and resolution of the identity operator

/A.6/ $\quad 1 = \int |\alpha\rangle\langle\alpha^*| \frac{d\alpha^* d\alpha}{\langle\alpha^*|\alpha\rangle}$.

For several cells the coherent states can be represented as tensor product of coherent states for one qbit cell

$$|\alpha\rangle = |\alpha_1\rangle \otimes \ldots \otimes |\alpha_N\rangle, \quad \langle\alpha^*| = \langle\alpha_N^*| \otimes \ldots \otimes \langle\alpha_1^*|.$$

We also use notations

$$|\alpha\rangle = |\alpha_1, \ldots, \alpha_N\rangle, \qquad \langle\alpha^*| = \langle\alpha_1^*, \ldots, \alpha_N^*|.$$

Let us consider matrix elements in coherent states (or in other terms matrix symbols) of operators $A$ and states $f$ on the space $F_N$

$$\langle\alpha^*|f, \quad \langle\alpha^*|A|\beta\rangle.$$

With use of resolution of the identity operator Eq. /A.6/ the matrix element of the state $Af$ can be written as

$$\langle\alpha^*|Af = \int \langle\alpha^*|A|\beta\rangle\langle\beta^*|f \frac{d\beta^* d\beta}{\langle\beta^*|\beta\rangle}.$$

For the product of operators

$$\langle\alpha^*|AB|\gamma\rangle = \int \langle\alpha^*|A|\beta\rangle\langle\beta^*|B|\gamma\rangle \frac{d\beta^* d\beta}{\langle\beta^*|\beta\rangle}.$$

It is also used normalized matrix symbols (sometimes called covariant symbols)

$$f(\alpha^*) = \frac{\langle\alpha^*|f\rangle}{\langle\alpha^*|0\rangle}, \quad A(\alpha^*, \beta) = \frac{\langle\alpha^*|A|\beta\rangle}{\langle\alpha^*|\beta\rangle}.$$

For the states both symbols coincide according to Eq. /A.5/.

For an operator action

$$Af(\alpha^*) = \int A(\alpha^*, \beta) f(\beta^*) \frac{\langle\alpha^*|\beta\rangle}{\langle\beta^*|\beta\rangle} d\beta^* d\beta.$$

For the product of operators

/A.7/ $\quad AB(\alpha^*, \gamma) = \int A(\alpha^*, \beta) B(\beta^*, \gamma) \frac{\langle\alpha^*|\beta\rangle\langle\beta^*|\gamma\rangle}{\langle\alpha^*|\gamma\rangle\langle\beta^*|\beta\rangle} d\beta^* d\beta.$

The right part of the last one relation defines so called convolution of symbols – ∘

$$AB(\alpha^*, \gamma) = A(\alpha^*, \beta) \circ B(\beta^*, \gamma).$$

Note that in contrast to the covariant symbols the matrix symbols can be defined for operators acting between different spaces

$$A: F_N \to F_M.$$

Grassmann algebra generators are often treated as variables of classical phase space of fermions system. So symbols are the elements of $\Lambda_{2N}$ i.e. functions on phase space and realize quantum to classical correspondence principle:

/A.8/ $\quad f \to f(\alpha^*),$
$\quad A \to A(\alpha^*, \beta),$
$\quad Af \to A(\alpha^*, \beta) \circ f(\beta^*),$
/A.9/ $\quad AB \to A(\alpha^*, \beta) \circ B(\beta^*, \gamma).$

For the element of dual space $g \in F_N^+$ the symbol is the element of $\Lambda_N$ defined as

$$g(\alpha) = \frac{g|\alpha\rangle}{\langle 0|\alpha\rangle}.$$

For $f, g \in F_N$ Hermitian scalar product is defined by

$$\langle f, g \rangle = f^+(g)$$

and its symbolic representation is

$$\langle f, g \rangle = \int f^+|\alpha\rangle\langle\alpha^*|g \frac{d\alpha^* d\alpha}{\langle \alpha^*|\alpha\rangle} = \int f(\alpha^*)^* g(\alpha^*) \frac{d\alpha^* d\alpha}{\langle \alpha^*|\alpha\rangle}.$$

Covariant symbols of creation and annihilation operations are

$$a_k^- \to \alpha_k, \quad a_k^+ \to \alpha_k^*.$$

Normal forms of operators and states are simply related with their covariant symbols by means of these maps [6]. Note that symbols of the states /A.8/ are the elements of subalgebra $\Lambda_N^*$ in Grassmann algebra $\Lambda_{2N}$ with half number of generators /A.3/. Generators of Grassmann subalgebra $\Lambda_N^*$ can be treated as variables of configuration space. As for bosons in case of fermions it can be given so called Fock-Bargmann representation of the algebra of creation and annihilation operations. Namely with /A.8/ we have identification

$$F_N \to \Lambda_N^*.$$

Creation operators are represented as above with left multiplication

$$a_k^+ \to \alpha_k^*,$$

and annihilation operators are represented with left partial derivative

$$a_k^- \to \frac{\partial}{\partial \alpha_k^*}.$$

Any operator $A$ on $F_N$ can be represented by creation and annihilation operations [6]. So in Fock-Bargmann representation operators are partial derivative operators $A\left(a^*, \frac{\partial}{\partial \alpha^*}\right)$ on Grassmann subalgebra $\Lambda_N^*$.

## References

1. A. M. Perelomov. Generalized Coherent States and Their Applications. Springer-Verlag (1986)
2. J. L. Martin. The Feynman principle for a Fermi system. Proc. Roy. Soc. Lond. A 251, n.1267. 543–549 (1959)
3. Y. Ohnuki, T. Kashiva. Coherent states of fermi operators and path integral. Progr. Theor. Phys. 60, n.2, 548–564 (1978)
4. C. Montonen. Multiloop amplitudes in additive dual-resonance models. Nuovo. Cim. 19A, n.1, 69–89 (1974)
5. I. Bars, M. Gunaydin. Unitary representations of non-compact supergroups Commun. Math. Phys. 91, n.1, 31–51 (1983)
6. F. A. Berezin. Feynman path-integrals in a phase-space. Sov. Phys. Usp. 5, 497–548 (1980)
7. T. Sleator, H. Weinfurter. Realizable Universal Quantum Logic Gates. Phys. Rev. Lett. 74, n. 20, 4087–4090 (1995)
8. A. Barenco, C. H. Bennett, R. Cleve, D. P. DiVincenzo, N. Margolus, P. Shor, T. Sleator, J. A. Smolin, H. Weinfurter. Phys. Rev. A 52, 3457 (1995)
9. A. Muthukrishnan. Classical and Quantum Logic Gates. An Introduction to Quantum Computing. Quantum Information Seminar, pp. 1–22, Rochester Center for Quantum Information (RCQI) (1999)
10. D. Deutsch. Quantum computational networks. Proc. Roy. Soc. Lond. A 425, 73–90 (1989)
11. S. Jafarpour. Introduction to the world of quantum computers. Proceedings of the Fifth IEEE International Conference on Cognitive Informatics (ICCI 2006) vols 1, 2 pp. 760–764 (2006)
12. N. da Costa, J. Kouneiher. Superlogic manifolds and geometric approach to quantum logic. International Journal of Geometric Methods in Modern Physics 13, 1550129 (2016)
13. F. A. Berezin. Introduction to Superanalysis. Mathematical Physics and Applied Mathematics, vol. 9, Springer (1987)
14. J. Kouneiher, A. P. M. Balan. Propositional manifolds and logical cohomology. Synthese 125, Issue 1, 147–154 (2000)
15. R. Guitart. The field $F_8$ as a Boolean manifold. Tbilisi Mathematical Journal 8(1), 31–62 (2015)
16. S. Eilenberg. Automata, languages and machines. Vol. B, Academic Press, New York, San Francisco, London (1976)
17. V. V. Smirnov. Feynman principle for fermions with odd covariant symbol of the Hamiltonian. Theor. and Math. Phys. 90, n.1, 102–106 (1992)
18. V. V. Smirnov. Path integral for system with spin. J. Phys. A . Math. and Gen. 32, n.7, 1285–1290 (1999)